# THE B-V AND V-R COLOR INDICES
## ON THE SURFACE OF NEA (214088) 2004 JN13

Albino Carbognani
Astronomical Observatory of the
Aosta Valley Autonomous Region (OAVdA)
Lignan 39, 11020 Nus (Aosta), ITALY
albino.carbognani@gmail.com

This paper presents the results of photometric observations with standard broad-band Bessel filters B, V, and R on near-Earth asteroid (214088) 2004 JN13. The analysis shows that the B-V and V-R color indices are fairly constant on the asteroid surface with mean values B-V = 0.83 ± 0.02 mag and V-R = 0.48 ± 0.03 mag, indicative of a relatively homogeneous surface color distribution.  For a typical albedo, assuming these colors indicate an S-type asteroid, a mean diameter of 2.4 ± 0.5 km is inferred.

The asteroid (214088) 2004 JN13 was discovered by LINEAR at Socorro on 2004 May 15 and came close to the Earth (about 0.137 AU) on 2014 Nov. 18. Based on the orbital parameters of $a = 2.8774$ AU, $e = 0.69728$, and $q = 0.871047$ AU (JPL, 2016), it appears that 2004 JN13 belongs to the Apollo class of near-Earth asteroids (NEA). Its rotation period is about 6.342 hours (Warner *et al.*, 2009). This object was observed from OAVdA after the Earth flyby to find the B-V and V-R color indices as a function of rotation phase to see if there were any color variations on the asteroid's surface that may indicate possible compositional inhomogeneity or differential space weathering process, as for NEA (297274) 1996 SK (Lin *et al.*, 2014).

In this paper, I will first discuss the instruments and tools used for the observations and the data reduction. Then a comparison will be made between the magnitudes reported in Landolt's reference catalog and those of APASS catalog for the same stars. This will show that the magnitudes of APASS catalog have an acceptable uncertainty compared to standard stars and that an atmospheric-instrumental model capable of transforming instrumental magnitude in standard magnitudes can be used to find the asteroid's true magnitude. Finally, I will look at the color indices of the asteroid and results that may be inferred.

### Instruments, Observations and Reduction Procedure

The asteroid (214088) 2004 JN13 was observed from OAVdA on 2014 Dec 16-17 from 19:00 UT to 02:30 UT, when it was moving away from the Earth but still bright, V ~ 13.9. The sky was clear and there were no passing clouds, so the transparency conditions were reasonably stable (see Fig. 1 and Fig. 2). The air mass values changed from 1.79 at the beginning of the observations, reached a minimum of 1.15, and increased to 1.64 at the end of observations. The images were captured with a modified 0.81-m *f*/7.9 Ritchey-Chrétien telescope and FLI-1001E CCD camera with an array of 1024×1024 pixels. The field-of-view was 13.1×13.1 arcmin while the plate scale was 1.54 arcsec/pixel in 2×2 binning mode.

Observations were performed alternately using broad-band Bessel B, V, and R filters with exposure times of 60 s for the B filter and 30 s for the others. The SNR of the target was greater than 100, which was ideal to obtain the color indices. The long observation run of 7.5 hours was divided into two sessions due to the proper motion of the asteroid, which was about 2.14 arcsec/min. This meant that different comparison stars were used for the first (18:55 to 23:45 UT) and second (23:50 to 02:30 UT) sessions (see Fig. 10). All images were calibrated with master-dark/bias and master-flat frames.

Reduction of the data, which consisted of the instrumental magnitudes of the target and comparison stars vs. the Julian day and air mass, and lightcurve analysis were done using *MPO Canopus* v10.7.1.3 (Warner, 2009), which makes use of differential aperture photometry and the Fourier period analysis algorithm developed by Harris (Harris *et al.*, 1989). The rotation period was found to be in good agreement with the known value (see Fig. 3-5).

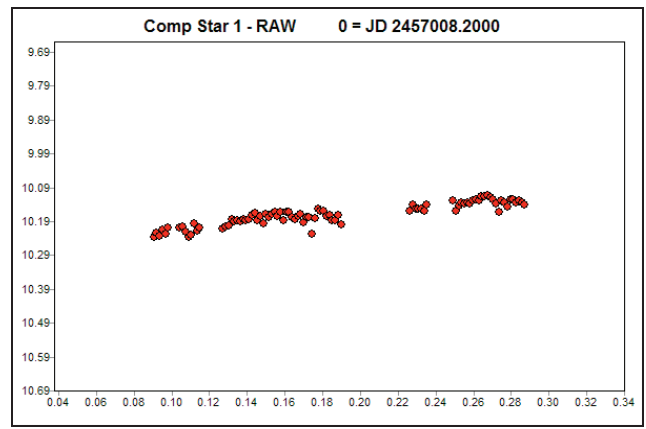

Figure 1. The raw magnitudes of the first comparison star for the first session shows no sudden or large attenuation indicating changing transparency. The gaps are when the asteroid was near background stars.

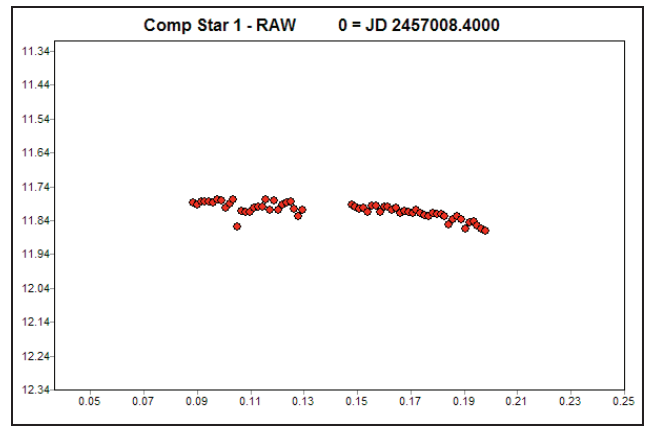

Figure 2. The raw magnitudes of the first comparison star for the second session also show no sudden attenuations. The gap is due to the asteroid being near a bright star.

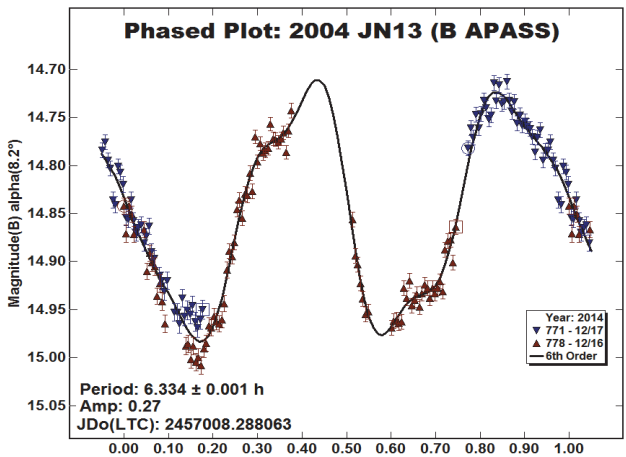

Figure 3. The phased lightcurve of 2004 JN13 taken with the B filter and reduced with B APASS magnitudes.

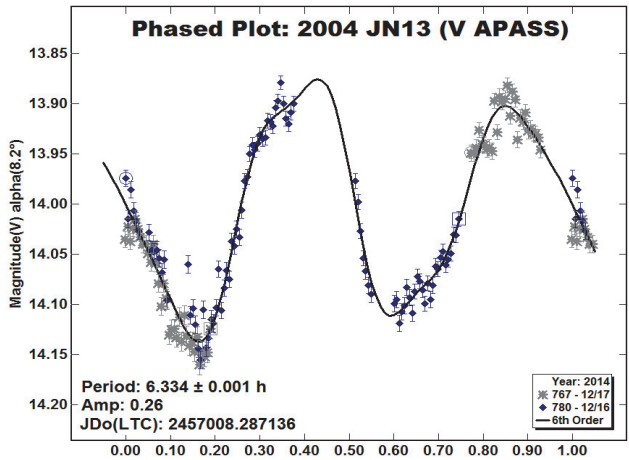

Figure 4. The phased lightcurve of 2004 JN13 taken with the V filter and reduced with V APASS magnitudes.

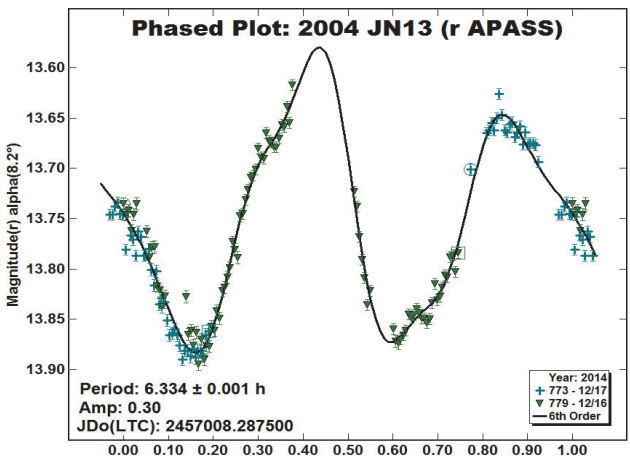

Figure 5. The phased lightcurve of 2004 JN13 taken with the R filter and reduced with r' (SR) APASS magnitudes.

### Landolt vs. APASS Star Catalog

In order to define the true magnitude of the asteroid in B, V, and R, comparison star magnitudes were taken from Release 9 of the AAVSO Photometric All-Sky Survey catalog (APASS; Henden *et al.*, 2009). This catalog contains photometry for 60 million objects over about 99% of the sky. The 5-band photometry is based on Johnson B and V and Sloan g' (SG), r' (SR), and i' (SI) filters. The catalog is a valid reference for stars over the range of V = 10-17 mag (Henden *et al.*, 2009). In this case however, stars with V > 15 were excluded because the photometric quality decreases significantly below this limit. The APASS catalog is not perfect, for example, Release 9 has known issues with blue magnitudes in the Northern Hemisphere and with red magnitudes in the Southern Hemisphere, and so it should be used with caution.

As seen in Fig. 6 and Fig. 7, for B and V between 11-15 mag, the RMS is about 0.04 mag when compared against Landolt's standard reference of 526 stars centered on the celestial equator (Landolt, 1992). So, for B and V

$$B_{Landolt} \equiv B_{APASS}$$
$$V_{Landolt} \equiv V_{APASS} \qquad (1)$$

Things are a bit different in the case of r' (SR) filter. In this case, there is a systematic shift of about 0.21-0.22 mag (Fig. 8). The equation to transform the SR mag in APASS catalog to a Landolt R mag is

$$R = SR - 0.112 - 0.128 \cdot (B - V) \qquad (2)$$

The RMS when using Eq. 2 is about 0.05 mag (Fig. 9).

In conclusion, if willing to accept a few hundredths of a mag decrease in accuracy and using Eq. 1 and 2, the stars from the APASS catalog in the same field as the asteroid can be used as references, provided that the night has constant atmospheric transparency conditions. Of course the APASS catalog is not a substitute of Landolt's fields in every situation; there may be cases in which a few hundredths of magnitude are important.

### The Selection of Comparison Stars

The comparison stars were selected with the *MPO Canopus* Comp Star Selector (CSS) utility using the appropriate B, V, or SR magnitude from the APASS catalog. This way, it was possible to select five non-variable comparisons stars in the same field as the asteroid, each having an SNR > 100, for use in differential photometry (Fig. 10).

The comparison stars are not necessarily the same in all the filters used, although there is a common subset (see Table I). These are the stars that were used to calibrate the instrumental-atmospheric local model that will be described in the following section. This mathematical model is able to transform the instrumental mag of the target to exoatmospheric (true) B, V and SR magnitudes.

| N | V | B | SR | RA | Dec | S |
|---|---|---|---|---|---|---|
| 1 | 12.603 | 13.814 | 12.152 | 04:59:56.32 | +16:46:17.3 | 1 |
| 2 | 13.144 | 14.480 | 12.633 | 05:00:02.57 | +16:54:10.3 | 1 |
| 3 | 12.275 | 13.079 | 11.998 | 05:00:12.57 | +17:07:26.4 | 2 |
| 4 | 12.930 | 13.752 | 12.640 | 04:59:29.97 | +17:06:31.6 | 2 |
| 5 | 13.299 | 13.973 | 13.072 | 04:59:47.05 | +17:07:40.0 | 2 |
| 6 | 14.507 | 15.282 | 14.246 | 05:00:04.60 | +17:01:59.0 | 2 |

Table I. The subset of comparison stars common to all filters that were used for the calibration of the instrumental-atmospheric local model (J2000.0 coordinates). The mag values are from APASS catalog while column 'S' indicates the sessions number. These stars are identified by their number (N) in Fig. 10.

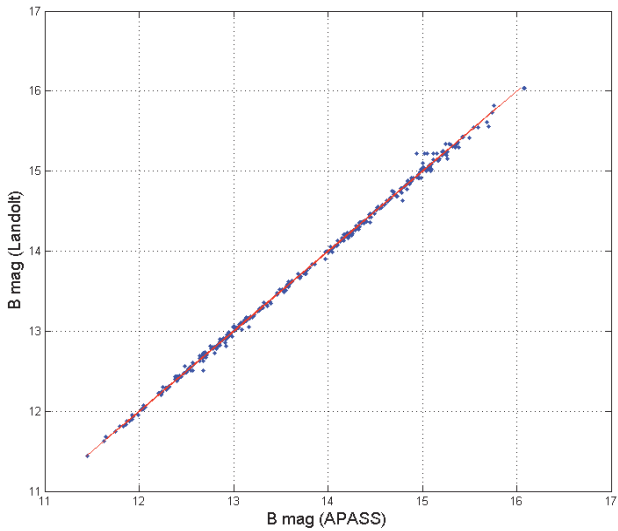

Figure 6. Landolt vs. APASS B magnitudes for the same stars. The red line is $B_{Landolt}$ vs. $B_{APASS}$. The RMS is about 0.04 mag.

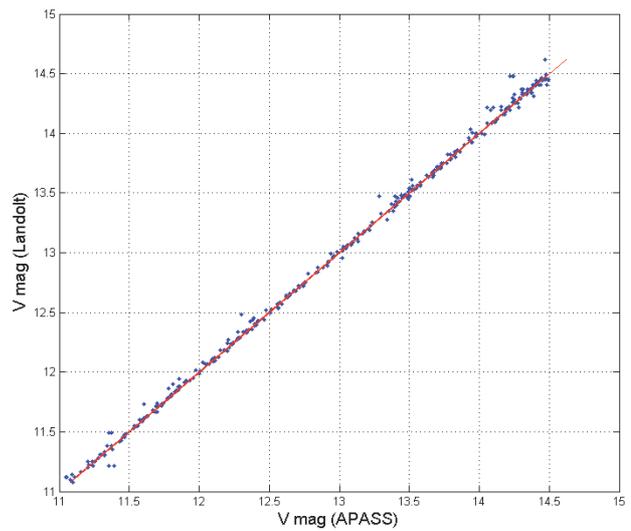

Figure 7. Landolt vs. APASS V magnitudes for the same stars. The red line is $V_{Landolt}$ vs. $V_{APASS}$. The RMS is about 0.04 mag.

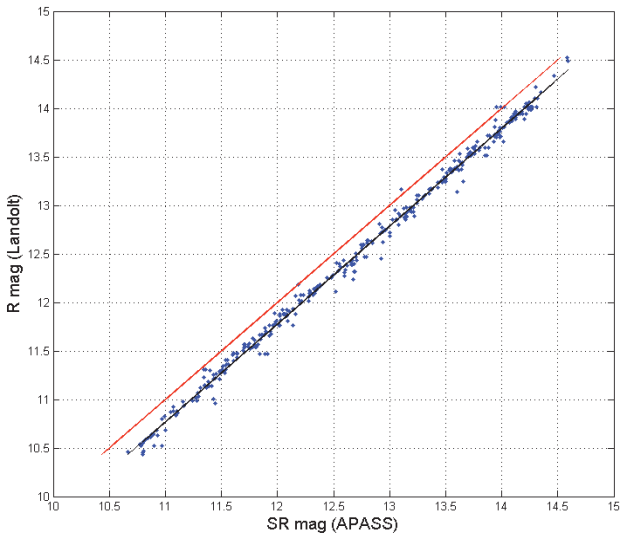

Figure 8. Landolt vs. APASS SR magnitudes for the same stars. The red line is $R_{Landolt}$ vs. $V_{APASS}$. The RMS is about 0.04 mag. There is a systematic shift of about 0.21-0.22 mag. The RMS of the fitted black line is about 0.07 mag.

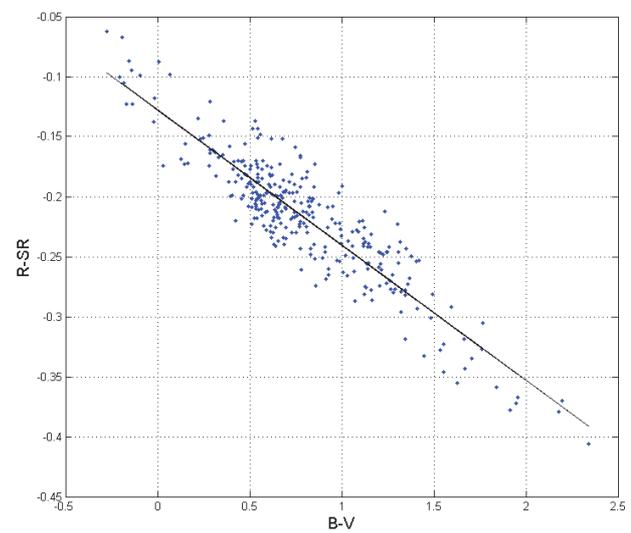

Figure 9. The R-SR mag vs. Landolt B-V. The RMS is about 0.05 mag.

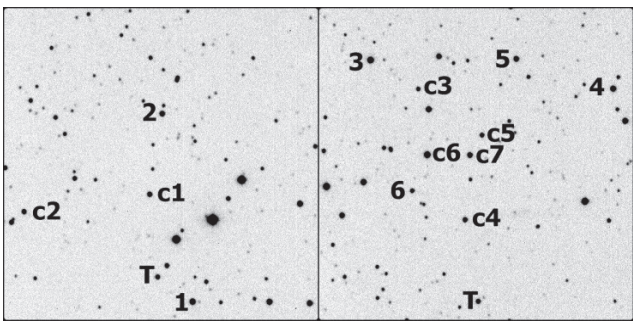

Figure 10. The fields for the first (left) and second (right) session showing the comparison stars in Table I along with 2004 JN13 (T) and the check stars (ci) listed in Table II. North is up, east is right.

## The Instrumental-Atmospheric Local Model

In order to derive the true B, V, and R magnitude of the target, it is necessary to create a model that defines the zero point of the mag scale and corrects for atmospheric absorption and instrumental color shift compared to standard magnitudes. This model should be valid only for the stars of the two fields, so it will be "local." The adopted equations for the instrumental-atmospheric local model using B-V are the same for all-sky photometry, i.e. (Harris *et al.*, 1981):

$$\begin{cases} B - b = Zb - \left(k_b' + k_b''(B-V)\right) \cdot X + C_b(B-V) \\ V - v = Zv - \left(k_v' + k_v''(B-V)\right) \cdot X + C_v(B-V) \end{cases} \quad (3)$$

The corresponding equations for V and SR are,

$$\begin{cases} SR - sr = Zsr - \left(k'_{sr} + k''_{sr}(V - SR)\right) \cdot X + C_{sr}(V - SR) \\ V - v = Zvr - \left(k'_{vr} + k''_{vr}(V - SR)\right) \cdot X + C_{vr}(V - SR) \end{cases} \quad (4)$$

In Eq. 3 and 4, B, V, and SR are taken from the APASS catalog. $Z_b$, $Z_v$, $Z_{sr}$, $Z_{vr}$ are the zero point magnitudes; b,v, and sr are the instrumental magnitude in the three filters; k′ is the first-order atmospheric extinction coefficient for the given filter; k′′ is the second-order atmospheric extinction; X is the air mass, and C is the instrumental color-correction coefficient for the given color index.

In general, nonlinear regression can be used to find a model based on a set of data points. MATLAB by MathWorks (*http://www.mathworks.com*) was used for this step.

To get the unknown coefficients, images of the six comparison stars from Table I were taken over a range of air masses ranged from X = 1.15 to 1.79. Eq. 3 and 4 can be used as two overdetermined linear systems (i.e., when there are more equations than unknowns) and an ordinary least squares method can be used to find an approximate solution.

Table II shows the various parameters of the instrumental-atmospheric local model using MATLAB.

| B | | V | | SR | | V | |
|---|---|---|---|---|---|---|---|
| Zb | 22.41 | Zv | 22.58 | Zsr | 22.80 | Zvr | 22.55 |
| $k_b$ | 0.195 | $k_v$ | 0.198 | $k_{sr}$ | 0.053 | $k_{vr}$ | 0.189 |
| $k_b''$ | 0.0013 | $k_v''$ | −0.062 | $k_{sr}''$ | 0.081 | $k_{vr}''$ | 0.053 |
| $C_b$ | 0.065 | $C_v$ | −0.177 | $C_{sr}$ | 0.196 | $C_{vr}$ | −0.415 |

Table II. The coefficients of the instrumental-atmospheric local model.

To check the results, the B, V, and SR magnitudes of a set of stars that were not used to compute the model (Table III) were computed and compared to the APASS values (Fig. 11 and 12).

| I | V | B | SR | RA | Dec | S |
|---|---|---|---|---|---|---|
| c1 | 13.549 | 15.168 | 12.951 | 05:00:04.34 | +16:50:45.3 | 1 |
| c2 | 13.793 | 14.696 | 13.458 | 05:00:26.24 | +16:49:51.9 | 1 |
| c3 | 14.612 | 15.658 | 14.250 | 05:00:04.07 | +17:06:16.9 | 2 |
| c4 | 13.722 | 15.367 | 13.084 | 04:59:55.19 | +17:00:49.5 | 2 |
| c5 | 14.164 | 15.238 | 13.764 | 04:59:52.67 | +17:04:24.8 | 2 |
| c6 | 12.231 | 13.506 | 11.758 | 05:00:02.15 | +17:03:31.4 | 2 |
| c7 | 13.301 | 14.224 | 12.961 | 04:59:54.69 | +17:03:33.6 | 2 |

Table III. The set of check stars shown in Fig. 10 used to test the quality of the instrumental-atmospheric local model (J2000.0 coordinates). The mag values are from APASS catalog while the 'S' column indicates the session number.

The Color Indices on the Asteroid's Surface

Now that the coefficients of the instrumental-atmospheric local model are known, it is possible to compute the (B-V) and (V-R) color indices for the target using its B, V, and R instrumental magnitudes. From Eq. (3) and Eq. (4) it follows that

$$(V - SR) = \frac{(v - sr) + X \cdot \left(k'_{sr} - k'_{vr}\right) + Zvr - Zsr}{1 - X \cdot \left(k''_{sr} - k''_{vr}\right) - \left(C_{vr} - C_{sr}\right)} \quad (5)$$

$$(B - V) = \frac{(b - v) + X \cdot \left(k'_v - k'_b\right) + Zb - Zv}{1 - X \cdot \left(k''_v - k''_b\right) - \left(C_b - C_v\right)} \quad (6)$$

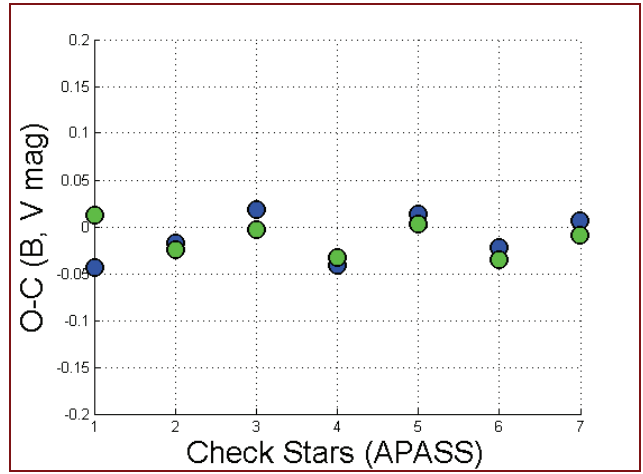

Figure 11. A graph showing the differences between the observed and the computed magnitudes for the check stars using the instrumental-atmospheric local model derived from data at air masses X = 1.19 and X = 1.79. Blue points are B and green points are V. The RMS in B is 0.026 mag; it is 0.018 mag in V.

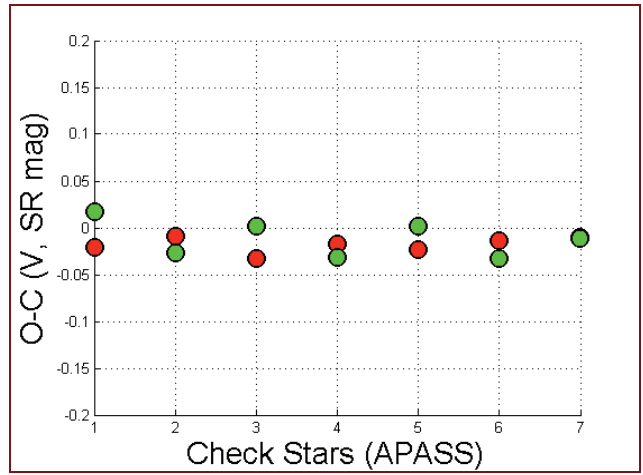

Figure 12. A graph showing the difference between the observed and the computed mag for the check stars using the derived instrumental-atmospheric model. Green points are V and red points are R. The RMS in R is 0.01 mag; it is 0.02 mag in V.

A correction is necessary to transform the color index V-SR to V-R. This can be found by subtracting Eq. 2 from the V mag,

$$(V - R) = (V - SR) + 0.112 + 0.128 \cdot (B - V) \quad (7)$$

Due to the propagation of errors, the RMS in Eq. 7 is about $\sqrt{0.04^2 + 0.05^2}$, or 0.06 mag. Using the B, V, and R instrumental magnitudes of the target in Eq. 5-7, the (B-V) and (V-R) color indices can be found as a function of the rotation phase. The uncertainty in a single measure is found by adding all the uncertainties in quadrature. For (B-V), this gives 0.07 mag and 0.09 mag for V-R. The final result is shown in Fig. 13.

Each red or blue point in Fig. 13 is the average of several black points obtained directly from the atmospheric-instrumental local model. The error bars indicate the corresponding standard deviation. The bin is 0.15 of the rotational phase, so there are seven "mean" points. As can be seen, the data suggest that there are no significant changes in either color index over a rotation.

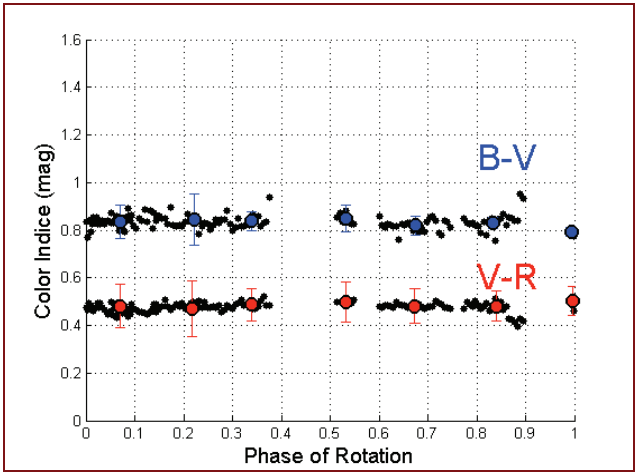

Figure 13. The mean (B-V) and (V-R) color indices of NEA (214088) 2004 JN13 vs. rotation phase. Black points are raw color indices. From rotation phase 0.0 to 0.2, the points are from both sessions. From 0.20 to 0.75, the points are from the second session only. From 0.75 to 1.0, they are from the first session only.

A Simpler Method for the Color Indices

A simplified method to achieve similar results follows. Looking at Eq. 5 and Eq. 6, the terms in the second member are all constants except for the difference between instrumental magnitudes, (v-sr) or (b-v), and the air mass, X. However, if the air mass changes little (from 1.15 to 1.64 in this case), Eq. 5 and 6 become

$$V - SR \cong a_1 + a_2 \cdot (v - sr) \qquad (8)$$

$$B - V \cong a_3 + a_4 \cdot (b - v) \qquad (9)$$

Where $a_i$ (i = 1, 2, 3, 4) are all constants. Figures 14 and 15 shows the calibration plot obtained with the comparison stars listed in Table I. Note that for each true color index, there are several instrumental color indices, with only a small variation, due to the small variation in air mass. Note also that the range of color indices of the comparison stars is sufficiently extended to determine the slope. In other words, not just solar-color stars should be used for the calibration. Using the available data and Eq. 8 and 9 gives

$$V - SR \cong -0.32 + 0.75 \cdot (v - sr) \qquad (10)$$

$$B - V \cong -0.18 + 1.17 \cdot (b - v) \qquad (11)$$

The RMS of Eq. 10 is 0.014 mag, while for Eq. 11 the RMS is 0.025 mag. At this point we can use the previous equations to switch between the instrumental color indices of the target to the true ones, making sure to include Eq. 7. The result is shown in Fig. 16, which is nearly the same as Fig. 13. In this case the uncertainty for the single black dots is 0.06 mag for (B-V) and 0.09 mag for (V-R). The error bars in Fig. 16 are the standard deviation of the mean values (red and blue dots).

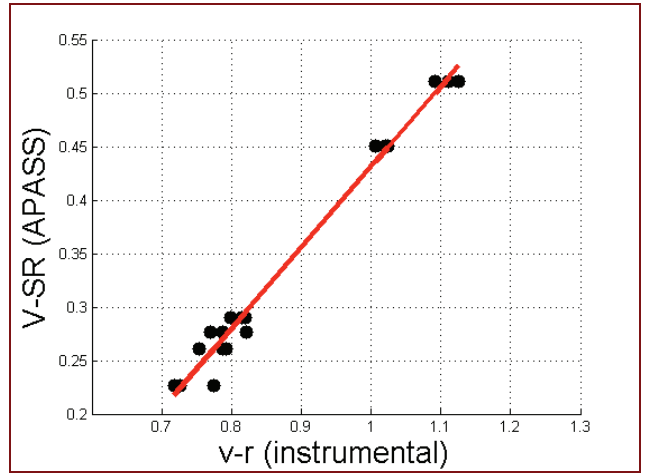

Figure 14. Plot of the catalog color index (V-SR) vs. the instrumental color index (v-r) for the comparison stars of Table I.

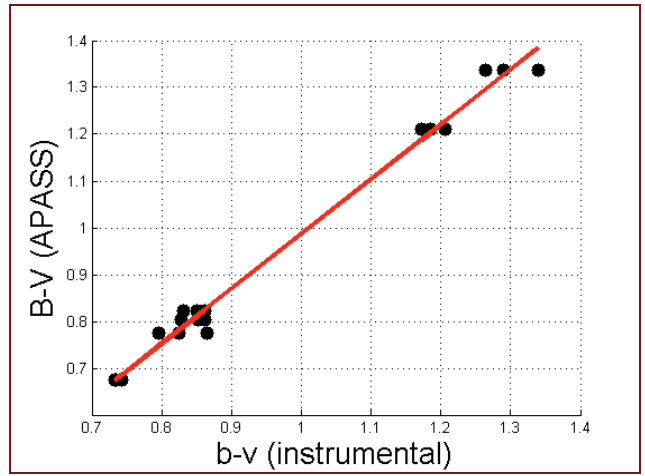

Figure 15. Plot of the catalog color index (B-V) vs. the instrumental color index (b-v) for the comparison stars of Table I.

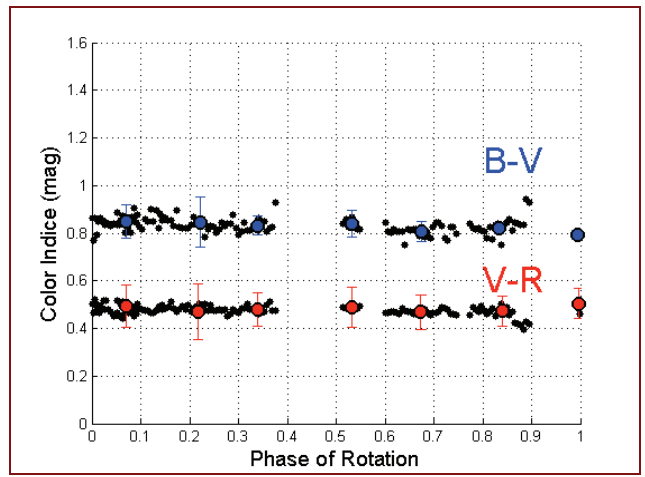

Figure 16. The color indices (B-V) and (V-R) of NEA (214088) 2004 JN13 vs. rotation phase using the simplified reduction method. This figure is very similar to Fig. 13.

The Optical Colors of NEAs and
the Taxonomic Class of 2004 JN13

From Fig. 13 or 16, the mean color indices values for 2004 JN13 are (B-V) = 0.83 ± 0.02 mag and (V-R) = 0.48 ± 0.03 mag.

The uncertainties are the standard deviation of the mean, so lower by a factor $1/\sqrt{7}$ = 0.38 with respect to the mean standard deviation of the points (red or blue). To find the class of the asteroid we can make a comparison with the data available in the scientific literature, where our choice is to use Dandy *et al.* (2003). Our mean (B-V) and (V-R) values most closely match the tabulated values for S-type asteroids; while Q- and V-types are not ruled out. Given the percentages of NEAs with a known taxonomic class, 52% belong to the S-type, 20% belong to the Q-type while only 7% belong to the V-type (Binzel, 2002). So it is more probable than not that 2004 JN13 it's an S-type asteroid.

Assuming that 2004 JN13 is an S-type asteroid and using the available data, a rough estimate can be made of asteroid's effective diameter. The mean absolute V magnitude in the H-G system is given by Bowell *et al.* (1989):

$$H_o = m_{v_r}(\alpha) + 2.5\log_{10}\left[(1-G)\Phi_1(\alpha) + G\Phi_2(\alpha)\right] \quad (12)$$

In Eq. 12, $H_o$ is the absolute magnitude at 0° phase angle (usually given as just 'H'), $m_{v_r}$ is the mean reduced magnitude and $\Phi_1$ and $\Phi_2$ are two functions of the phase angle $\alpha$. For an S-type asteroid, in the H-G system, G = 0.24 ± 0.11 (Shevchenko and Lupishko, 1998).

Substituting the observed phase angle of 8.2° and mean V magnitude of 14.01 ± 0.03 mag gives H = 15.47 ± 0.38 mag, which is in good agreement with the JPL Small-Body Database value of 15.3. Assuming that the asteroid is an S-type and assuming the mean geometric albedo for the class of $p_V$ = 0.20 ± 0.05, the effective diameter of 2004 JN13 will be (Harris, 1997)

$$D_e = \frac{1329}{\sqrt{P_v}}10^{-0.2H_V} = 2.4 \pm 0.5 \text{ km} \quad (13)$$

Conclusions

Asteroid (214088) 2004 JN13 was observed from OAVdA on 2014 Dec 16-17 for about 7.5 hours with B, V, and R filters on a night with relatively stable sky transparency. Using the stars from the APASS catalog as a reference, an instrumental-atmospheric local model was constructed. With this, it was possible to compute the color indices (B-V) and (V-R) as a function of the rotation phase. The color indices are constant within the uncertainties, indicating that the asteroid's surface is homogeneous within the measurement limits. The same results were achieved with a simplified data reduction method. The photometric data collected indicate that the asteroid is probably an S-type. Using a series of additional assumptions, the estimated effective diameter was found to be 2.4 ± 0.5 km.

Acknowledgements

This research has made use of the NASA's Astrophysics Data System and Planetary Data System, JPL's Small-Body Database Browser, and the VizieR catalogue access tool, CDS, Strasbourg, France. Research at the Astronomical Observatory of the Aosta Valley Autonomous Region was supported by the 2013 Shoemaker NEO Grant from The Planetary Society.